\documentclass[twocolumn,aps,floatfix,showpacs]{revtex4}
\newcommand{\be}{\begin{equation}}
\newcommand{\ee}{\end{equation}}
\newcommand{\bea}{\begin{eqnarray}}
\newcommand{\eea}{\end{eqnarray}}

\usepackage{graphicx}
\usepackage{bm}
\begin{document}

\title{Accurate near-threshold model for ultracold
 KRb dimers from interisotope Feshbach spectroscopy}

\author{Andrea Simoni } \affiliation{Institut de Physique de Rennes,
\\ UMR 6251 du CNRS and Universit\'e de
Rennes, 35042 Rennes Cedex, France}

\author{Matteo Zaccanti$^{1}$, Chiara D'Errico$^{1}$, Marco Fattori$^{1,2}$, Giacomo Roati$^{1}$,
Massimo Inguscio$^{1}$, and Giovanni Modugno$^{1}$} \affiliation{$^{1}$LENS and Dipartimento
di Fisica, Universit\`a di Firenze, and INFM-CNR\\  Via Nello Carrara 1,
50019 Sesto Fiorentino, Italy\\
$^{2}$ Museo Storico della Fisica e Centro Studi e Ricerche ''Enrico Fermi'',Compendio del Viminale, 00184 Roma, Italy}

\begin{abstract}

We investigate magnetic Feshbach resonances in two different ultracold
K-Rb mixtures. Information on the $^{39}$K-$^{87}$Rb isotopic pair is
combined with novel and pre-existing observations of resonance patterns
for $^{40}$K-$^{87}$Rb. Interisotope resonance spectroscopy improves
significantly our near-threshold model for scattering and bound-state
calculations. Our analysis determines the number of bound states
in singlet/triplet potentials and establishes precisely near threshold
parameters for all K-Rb pairs of interest for experiments with both
atoms and molecules. In addition, the model verifies the validity of
the Born-Oppenheimer approximation at the present level of accuracy.

\end{abstract}

\date{\today}
\maketitle

\section{introduction}

Magnetic Feshbach resonances \cite{feshbach, sodium} represent
a unique tool for manipulating atomic quantum gases: they allow one to
explore new regimes of strong interaction by modifying the collisional
properties in Bose gases \cite{Feshmag}, Fermi gases \cite{fermioni} and
mixtures \cite{mix, zaccanti, bongs}; they also enable the production of
ultracold weakly-bound molecules by means of magnetic field sweeps across
resonance both in homonuclear \cite{molecular gases} and heteronuclear
\cite{eteromol} systems. Moreover, the tight constraints set by Feshbach
spectroscopy on the position of molecular energy levels closest to
dissociation \cite{Chu, rempe} can lead to a very accurate determination
of long range interaction potentials and scattering properties of the
atomic system of interest.

A system that has attracted considerable interest is K-Rb: in fact
this mixture has several isotopic pairs that are easy to bring into
ultracold and quantum degenerate regimes \cite{modugno, roati2,
roati07, desarlo}, the main isotopic combinations present several
accessible Feshbach resonances \cite{ferlaino, KRbJila, KRblast},
and the ground state dimer has a relatively large electric dipole
moment \cite{kotochigova}.  Knowledge of molecular KRb potentials is
crucial for studying quantitatively most phenomena in this system: on
one side scattering lengths and dispersion coefficients are relevant
for characterizing atomic collisions and weakly bound dimers. On the
other side the short range potential well must be determined in order to
perform experiments with deeply bound molecules. The molecular potentials
of KRb have been so far constructed using different experimental inputs:
Fourier transform spectroscopy and photoassociation techniques, reported
most recently in Refs. \cite{stwalley, pashov}, lead to a detailed
knowledge of the short range potential behavior; Feshbach spectroscopy of
the $^{40}$K-$^{87}$Rb fermion-boson mixture \cite{ferlaino, KRbJila,
KRblast} has allowed the long range parameters of the system to be
determined very precisely. 

In this work we combine for the first time Feshbach spectroscopy on
two different isotopic pairs of K-Rb and show that this significantly
improves the threshold model precision. Moreover, the number of bound
states supported by the interaction potentials is univocally fixed, in
agreement with the recent values derived from molecular spectroscopy
\cite{pashov,pavel}. We can therefore use the model for different
isotopes without being limited by typical few bound states uncertainties
\cite{ferlaino,zemke}. Finally, availability of accurate interisotope
data allows us to test possible deviations from the Born-Oppenheimer
approximation.

The paper is organized as follows: section \ref{sec2} presents the
experimental procedure used to produce an ultracold sample and to
detect magnetic Feshbach resonances and zero crossings ({\it i.e.} the field
locations of vanishing scattering length). Section \ref{sec3} introduces
the theoretical model and is devoted to data and error analysis; near
threshold molecular levels for selected K-Rb isotopic pairs are also
presented. A brief conclusive discussion ends this work.

\section{Experimental methods}
\label{sec2}
In our experimental apparatus we have
investigated both the fermion boson $^{40}$K-$^{87}$Rb and the boson boson
$^{39}$K-$^{87}$Rb mixture. With respect to the techniques for the
realization and for Feshbach spectroscopy of the former mixture we
refer to \cite{ferlaino}, while we will focus here on the experimental
procedure concerning the $^{39}$K-$^{87}$Rb mixture. The apparatus and
techniques we used are similar to the ones we developed for the other
isotopomer \cite{ferlaino}, and have already been presented elsewhere
\cite{roati07}. In summary, we start by preparing a mixture of $^{39}$K
and $^{87}$Rb atoms in a magneto-optical trap at temperatures of
the order of few 100 $\mu$K. We simultaneously load the two species
in a magnetic potential in their stretched Zeeman states $|f_a=2,
m_{fa}=2\rangle$ and  $|f_b=2, m_{fb}=2\rangle$, and perform 25 s of
selective evaporation of rubidium on the hyperfine transition at 6.834
GHz. Potassium is sympathetically cooled through interspecies collisions
\cite{desarlo}. When the binary gas temperature is around 800 nK we
transfer the mixture in an optical potential. This is created by two
focused laser beams at a wavelength $\lambda$=1030 nm with beam waists
of about 100 $\mu$m, crossing in the horizontal plane.

In this work we have studied the ground state manifold  $f_{a,b}$=1 of the
$^{39}$K-$^{87}$Rb system. In general, Feshbach resonances can occur in
several mixtures of Zeeman sublevels; however, not all of these are stable
against spin-exchange inelastic processes.  Such processes conserve the
projection of the hyperfine angular momentum in the direction of the
magnetic field, $m_f=m_{fa}+m_{fb}$ and the orbital angular momentum
$\vec{\ell}$ of the atoms about their center of mass. If states having
internal energy lower than the initial one and the same value of $m_f$
exist, the system will in general undergo rapid spin-exchange decay.

\begin{figure}[htbp] \includegraphics[width=\columnwidth,clip]{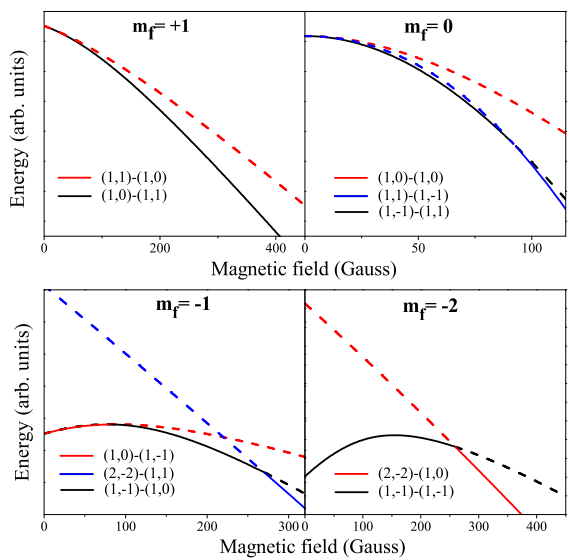}
\caption{Level scheme of different spin mixtures of
$^{39}$K and $^{87}$Rb with $m_f= +1, 0 , -1, -2$. For each spin
combination, the solid (dotted) line indicates the region where the
mixture is stable (unstable) against spin-exchange collisions. For each
$m_f$ only the magnetic field region where the levels
behavior is not trivial has been shown.}
\label{fig1}
\end{figure}

In Fig.\ref{fig1} the energies of different combinations of Zeeman states
identified by the value of $m_f$ are shown, and the stability region of
every mixture is marked with a solid line. As convention, the first state
refers to potassium and the second one to rubidium.

We have investigated the combinations $| 1, 1 \rangle+| 1, 1
\rangle$ and $| 1, 0 \rangle+| 1, 1 \rangle$, that are always stable, and $|
1, -1 \rangle+| 1, -1 \rangle$ in its stability region. The atoms are
initially prepared in $| 1, 1 \rangle$ by two consecutive adiabatic rapid
passages over the hyperfine transitions around 485 MHz for $^{39}$K
and 6857 MHz for $^{87}$Rb, in a 10 G homogeneous magnetic field. The
transfer efficiency is typically better than 90 percent and the non
transferred atoms are removed by means of a few microsecond
blast of resonant light. Both species are transferred from $|1, 1\rangle$
to $|1, -1\rangle$ state by applying a radio frequency sweep about
7.6 MHz at a 10 G field. For transferring potassium atoms from the $|1,
1\rangle$ to the $|1, 0\rangle$ state we ramp the magnetic field up to 38.5
G, where the Zeeman splitting of potassium and rubidium already differ
by some MHz, and apply a radio frequency sweep around 28.5 MHz. Once the
desired mixture is prepared we change the external magnetic field in few
tens of ms and actively stabilize it to any value below 1000 G, with a
short term stability of $\sim$ 30 mG and a long term one (day to day)
better than 100 mG. We calibrate the field by means of microwave and
radio frequency spectroscopy on two different hyperfine transitions of Rb.

Heteronuclear Feshbach resonances are detected as an enhancement of
three-body losses. In fact, the $s$-wave scattering length in the
vicinity of a resonance varies according to the dispersive behavior
\begin{equation}
a(B) = a_{\rm bg} \left( 1-\frac{\Delta}{B-B_{0}}
\right)
\end{equation}
where $a_{\rm bg}$ is the background scattering length, $\Delta$ is
the width of the resonance, defined as the distance between the zero
crossing and the resonance center $B_{0}$. As the scattering length
$a(B)$ diverges three body inelastic rates are enhanced ~\cite{Esry}
resulting in atom loss from the trap and heating.

We have at first searched several of the broadest resonances theoretically
predicted by the model of Ref. \cite{ferlaino}, which employed however
a number of singlet bound states two units smaller than the correct one
(see below); experimentally, all of them were found within a few Gauss
from the predicted positions. In general, the experimental location of
broader resonances is affected by larger uncertainties: consequently,
narrow features are crucial to improve the model precision as their
position can be determined with high accuracy. For an accurate
detection of such weak features, as for example the resonances near
248 G in the $|1, 1\rangle$+$|1, 1\rangle$ mixture or the one at 674
G in $|1, 0\rangle$+$|1, 1\rangle$ collisions (see Tab.~\ref{tab2})
we have performed further studies at lower temperatures (250-350 nK)
and higher densities.

Obtention of such conditions is crucial in particular for revealing
$p$-wave resonances, whose complex structure \cite{bohn} can be
easily masked by thermal effects.
The mixture is cooled by reducing the trap depth in $2.4$
seconds with an exponential ramp. The optical potential is designed in
such a way to force evaporation of rubidium along the vertical direction,
while potassium is sympathetically cooled without significant atom losses.
\begin{figure}[htbp] \includegraphics[width=\columnwidth,clip]{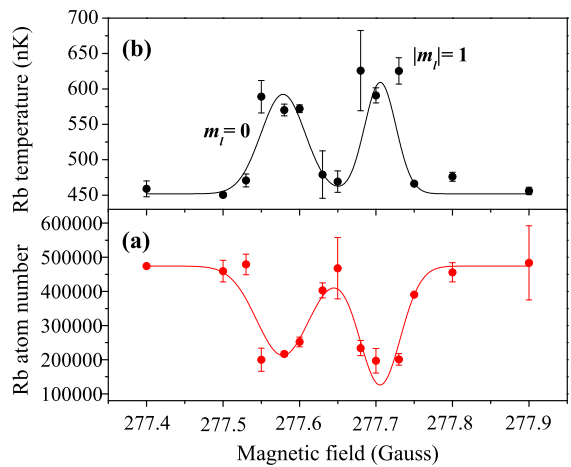}
\caption{(a): Experimental Rb loss features due to a $^{39}$K-$^{87}$Rb
heteronuclear $|1, 1\rangle + |1, 1\rangle$ $p$-wave resonance.
The doublet structure is peculiar of the resonance $p$-wave character.
Losses are accompanied by heating of the sample, as shown in panel (b). A
similar behavior has been observed on the potassium cloud. Gaussian profiles 
are fitted to the experimental data in order to extract the resonance centers.
The $|m_l|$ assignment of the two features is derived from a global fit of 
all the resonances (see text).} \label{fig2}
\end{figure}
As already remarked for $p$-wave scattering between fermions \cite{bohn}
and more recently in a $^{40}$K-$^{87}$Rb fermion boson mixture
\cite{bongs}, a doublet splitting represents direct evidence of the
$p$-wave character of such resonances: this feature arises from spin-spin
and second order spin-orbit interactions, as we will discuss later.
The typical doublet structure has been observed for two $p$-wave
resonances at 277.5 G (see Fig.~\ref{fig2}) and 495.5 G.
\begin{figure}[htbp]
\includegraphics[width=\columnwidth,clip]{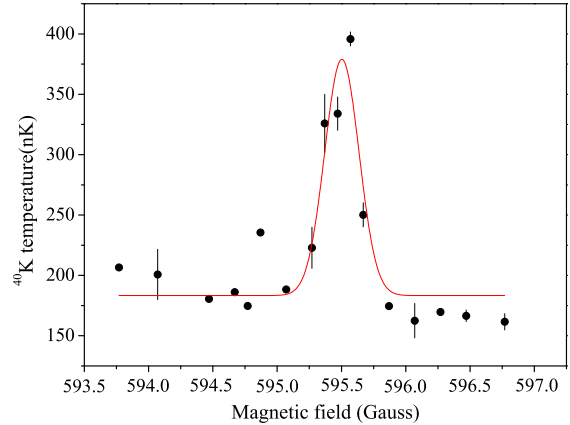}
\caption{ Experimental observation of the zero crossing near the 598.4 G
Fermi Bose resonance. As the magnetic field assumes values close to the
zero crossing point, the heterouclear elastic cross section vanishes
and the efficiency of sympathetic cooling during further evaporation
in the optical trap decreases. A similar behavior has been observed in
the Bose Bose mixture for the zero crossing associated to the resonance
located at 318 G, see Tab.~\ref{tab2}}
\label{fig3}
\end{figure}

The location of the zero crossing associated to broad Feshbach resonances
has been determined both in the Fermi Bose and Bose Bose mixtures
by recording the efficiency of sympathetic cooling of potassium as a
function of the magnetic field applied during the evaporation in the
optical potential, see \cite{zaccanti}. In fact, in the ultracold regime
the total elastic cross section vanishes with the $s$-wave scattering
length resulting in a strongly reduced sympathetic cooling rate. For this
kind of analysis we lower the beam intensities of the optical trap
and after 2.4 s of forced evaporation we measure the temperature of the
potassium cloud as a function of the external magnetic field. As shown
in Fig.~\ref{fig3}, the position of the zero crossing appears then as
a sharp peak in the potassium temperature.

\begin{table*}[t] \begin{center} \caption{Experimentally observed
magnetic-field positions $B_{\rm exp}$ and theoretically calculated
positions $B_{\rm th}$ for collisions of $^{39}$K and $^{87}$Rb. Few
experimental zero-crossing positions have also been used for modeling and
are identified with a data superscript $^*$.  Zeeman states of the atomic
fragments correlate in zero field with $| f_a m_{fa} \rangle$ and $|f_b
m_{fb} \rangle$, respectively (first column). Calculations use parameters
of Eq.~\ref{bestfit}. Errors shown in parenthesis represent one standard
deviation for both experimental and theoretical values. The magnetic
widths $\Delta$ are provided for the observed $s$-wave features. In view
of possible experiments of molecule formation the background scattering
length $a_{\rm bg}$ and magnetic moment $s$ are also given for resonances
in the lowest Zeeman sublevel.  The $\ell$ quantum number is the
orbital angular momentum of the molecule associated with each resonance.
The magnitude $|m|$ of its projection on the magnetic field is shown for
the $\ell =1$ doublet features that have been experimentally resolved.
Last column shows the spin coupling scheme of the Feshbach molecule
$\{t f_b f \}$, see text.
}  \vskip 12pt
 \label{tab2}
\begin{tabular}{ l | r@{.}l  r@{.}l   l c c c c}
\hline \hline
 $| f_a m_{fa}\rangle +| f_b m_{fb} \rangle $ & \multicolumn{2}{l}{$B_{\rm exp}$(G)}
 & \multicolumn{2}{l}{$B_{\rm th}$(G)} & $\Delta_{\rm th}$(G) & $a_{\rm bg}$~($a_0$) & $-s (\mu_B)$ & $\ell~(|m|)$ & assignment  \\
\hline
$ |1,1 \rangle + | 1,1 \rangle $  & 247&9(2) &  248&05(3)                 & 0.28 & 34 & 2.8 & 0  &  $\{ \frac{5}{2} 2 3  \}$  \\   %
                                       & 277&57(5) &  277&53(3)           &      &  & & 1(0)        &   \\
                                       & 277&70(5) &  277&70(3)           &      &  && 1(1)         &   \\
                                       & 317&9(5) & 318&30(3)             & 7.6  & 34 &2.0 & 0   &  $\{ \frac{5}{2} 2 2  \}$    \\   %
                                       & 325&4(5)$^*$&  325&92(3)$^*$     &      &  & & 0        &    \\      %
                                       & 495&19(6) &  495&19(3)           &      &  &&   1(0)       &    \\
                                       & 495&62(6)  & 495&65(3)           &      &  && 1(1)         &     \\
                                       & 531&2(3)  & 530&72(3)            & 2.5  & 35  &2.0 & 0  &   $\{  \frac{3}{2}  1 3 \}$    \\   %
                                       & 616&06(10)  & 615&85(4)          & 9.5[-2] & 35 & 1.9 & 0 &   $\{ \sim \frac{3}{2}1   2   \}$   \\    %
$ | 1 ,0  \rangle + | 1,1 \rangle $ & 623&47(6)  & 623&48(5)              & 6[-3]    & & & 0       &\\
                                       & 673&62(8)  &  673&76(4)          & 0.25     & & & 0      &\\
$ | 1 ,-1  \rangle + | 1,-1 \rangle $ & 117&6(4)  & 117&59(3)             & -1.3     & & & 0      & \\
\hline \hline
\end{tabular}
\end{center}
\end{table*}

\begin{table*}[t]
\begin{center}
\caption{
Same as Tab.~\ref{tab2} but for collisions of $^{40}$K and $^{87}$Rb. As
in Tab.~\ref{tab2}, few experimental zero crossing positions have been
used for theoretical modeling and are identified by a $^*$ symbol. Please
note that the quantity -$\Delta_{\rm th}$ is reported here. Last column
shows the spin coupling scheme of the Feshbach molecule $(f_a f_b f)$,
see text.}
\vskip 12pt
\label{labtb1}
\begin{tabular}{ l | r@{.}l  r@{.}l  l c c c c}
\hline \hline
 $| f_a m_{fa}\rangle +| f_b m_{fb} \rangle $ & \multicolumn{2}{l}{$B_{\rm exp}$(G)} & \multicolumn{2}{l}{$B_{\rm th}$(G)}
 &  -$\Delta_{\rm th}$(G)   &  $a_{\rm bg}$~($a_0$)   & $-s (\mu_B)$    & $\ell$ & assignment  \\
\hline
$ | 9/2,-9/2 \rangle + | 1,1 \rangle $  & 456&1(2) &  456&31(7)         & & &       & 1  \\
                                       & 495&6(5) &  495&31(12)        & 0.15  & -177  & 2.7  & 0 & $(\frac{7}{2} 1 \frac{7}{2})$\\
                                       & 515&7(5) &  515&35(7)         & & &       & 1 & \\
                                       & 543&3(5)$^*$ & 543&66(8)$^*$  & & &       & 0 &\\
                                       & 546&6(2)&  546&75(6)          & 3.1   & -189  & 2.3 & 0 & $(\frac{7}{2} 1 \frac{9}{2} )$  \\
                                       & 658&9(6) &  659&02(13)        & 0.80  & -196  & 2.8 & 0 & $(\frac{7}{2} 2 \frac{11}{2} )$  \\
                                       & 663&7(2)  & 663&80(10)        & & &       & 2 & \\
$ | 9/2,-7/2 \rangle + | 1,1 \rangle $  & 469&2(4)  & 469&03(13)        & 0.28 & & & 0 &  \\
                                       & 584&0(10)   &  584&01(11)     & 0.70  & & & 0 & \\
                                       & 591&0(3) &  590&85(7)         & & & & 2 & \\
                                       & 595&5(5)$^*$ &   595&60(7)$^*$& & & & 0 & \\
                                       & 598&4(2)  &  598&17(6)        & 2.53  & & & 0 & \\
                                       & 697&3(3) & 697&37(9)          & 0.15 & & & 0 & \\
                                       & 705&0(14) & 704&33(13)        & 0.82  & & & 0 & \\
$ | 9/2,-9/2 \rangle + | 1,0 \rangle $  & 542&5(5)$^*$  & 542&79(5)$^*$ & & & & 0  & \\
                                       & 545&9(2)    &  545&95(7)      & 3.2 & & & 0  &\\
                                       & 957&6(5)$^*$  & 957&70(13)$^*$& & & & 0 & \\
                                       & 962&1(2)  & 962&04(13)        & 4.3 & & & 0 & \\
$ | 9/2,7/2 \rangle + | 1,1 \rangle $ & 299&1(3)  & 298&51(5)         & 0.61  & & & 0  & \\
                                       & 852&4(8)  &  851&93(14)       & 6.1[-2] & & & 0  &\\
\hline \hline
\end{tabular}
\end{center}
\end{table*}

The magnetic field position of all observed Feshbach resonances is
reported in Tab.~\ref{tab2} and Tab.~\ref{labtb1}. We also report zero
crossing positions for few broad resonances and the doublet splitting
of $p$-wave resonances for the boson-boson mixture. Thirteen of the
boson-fermion features are from Ref.~\cite{ferlaino}.

\section{Theoretical analysis}
\label{sec3}

The main features of our theoretical model have already been described in
Ref.~\cite{ferlaino}. At variance with our previous work we adopt here
the spectroscopic singlet $^1\Sigma^+$ potential of Amiot~\cite{amiot}.
This potential supports the correct number $N^b_S$(40-87)$=100$ of
rotationless vibrational levels (see below) whereas the formerly used
Rousseau's {\it ab-initio} potential energy curve~\cite{rousseau} only
has 98. This difference is immaterial as far as one is concerned with
the study of a single isotope but it would be responsible for systematic
errors when properties of other pairs are calculated.

The singlet potential energy curve is obtained at regular internuclear
distances using the near-dissociation coefficients of~\cite{amiot}
and the RKR1 code~\cite{rkr1}. The triplet {\it ab-initio} potential
$^3\Sigma^+$ of Rousseau provides the correct number $N^b_T$(40-87)$=32$
of rotationless vibrational levels (see below) and is retained for our
analysis. We have now sufficient experimental information to determine
both leading long-range coefficients $C_6$ and $C_8$ independently of
{\it ab-initio} calculations. The model is also parameterized in terms
of $s$-wave singlet-triplet scattering lengths $a_{S,T}$ of the Fermi
Bose mixed system and includes relativistic spin-spin and second-order
spin order corrections~\cite{Mies}.

Our dataset comprises resonances observed in two isotopic mixtures.
The $^{40}$K-$^{87}$Rb fermion boson system is now well characterized
and theoretically understood~\cite{ferlaino,KRblast}. For the
$^{39}$K-$^{87}$Rb boson boson pair, the former theoretical predictions
of Ref.~\cite{ferlaino} is in good agreement with the present
observations. This circumstance is in itself sufficient to conclude that
the $N^b_T=32$ is correct, as a $\pm1$ variation in $N^b_T$ gives rise
to shifts of $^{39}$K-$^{87}$Rb Feshbach resonances as large as 10~G,
for fixed values of $a_{S,T}(40-87)$.

Shifts are in general less dramatic upon variation of the dissociation
energy of the deeper $^1\Sigma^+$ potential. In addition the boson boson
resonances observed here have mostly triplet character. Fortunately
the specific feature at $\sim 616$~G has sufficient singlet mixing
for its position to shift of about $\pm$3~G per bound state added
or subtracted from the $^1\Sigma^+$. This is sufficient to fix
conclusively $N^b_S=100$. Our present values of $N^b_{S,T}$ confirm
recent spectroscopic ~\cite{pashov} and {\it ab-initio} potentials
~\cite{private}.
After this preliminary characterization of the interaction potentials,
we proceed to fine-tune the potential shape in order to reproduce the
present experimental spectra. The presence of narrow resonances in our
data is crucial to improve the parameters precision, as their position
can be determined experimentally more accurately~\cite{KNJP}.

It has been recently remarked in Ref.~\cite{pashov} that the splitting
observed in the $^{40}$K-$^{87}$Rb mixture ~\cite{bongs} between $\ell=1$
resonances with different projections $m=0$ and $|m|=1$ of $\vec{\ell}$
along the magnetic field cannot be accounted for by electron spin-spin
interactions only. Comparison of two $\ell=1$ doublet features of the
boson boson system (see Tab.~\ref{labtb1} and Fig. \ref{fig2}) with
theoretical calculations confirms this observation. In particular,
the spin-spin induced splitting of the doublet at $495$~G is found
theoretically to be of $\sim$900~mG versus an observed value of
$\sim$500~mG. Within the present resolution
 this discrepancy might be sufficient to bias our analysis. Hence,
we perform a preliminary $\chi^2$ minimization only based on $s$-wave
resonances, which are virtually unaffected by spin interactions. Next,
we introduce a phenomenological second order spin-orbit operator of
the form~\cite{Mies}
\be
V_{so}(R)=\frac{C \alpha^2}{2}   e^{-\beta (R-R_S)}\left( 3 S_z^2 -S^2 \right)
\ee
where $\alpha$ is the fine-structure constant, $S$ is the total electrons
spin and $z$ is taken along the internuclear axis. We assign to the
parameters $\beta$ and $R_S$ the arbitrary yet reasonable values
$0.085 a_0^{-1}$ and $10 a_0$ whereas the strength $C$ is fixed to
$1.9 10^{-3}E_{\rm h}$ in order to reproduce the observed doublet
separations. We perform a final optimization including all $\ell >0$
features for both isotopes. Result of the fit is :
\bea
 a_S(40-87)&=&-110.6(4)\, a_0  \nonumber \\
 a_T(40-87)&=& -214.0(4)\, a_0  \nonumber \\
 C_6&=&4290(2)\, a_0^6 E_{\rm h} \nonumber \\
 C_8&=&4.79(4) 10^5 \, a_0^8 E_{\rm h} .
\label{bestfit}
\eea
The reduced chi-square (i.e. the $\chi^2$ per degree of freedom) is
$\tilde{\chi}^2=0.84$ and the maximum discrepancy with the empirical data
is less than two standard deviations. Note that the positions of $\ell
=2$ features, which are also shifted by spin interactions, are well
reproduced thus conferming the quality of our analysis. Our $a_{S,T}$
fully agree with the determination of Ref.~\cite{ferlaino}. The van der
Waals coefficient $C_6$ is consistent to about one standard deviation with
the value $4274(13)a_0^6 E_{\rm h}$ given by Derevianko {\it et al.} \cite{derevianko}
while $C_8$ deviates by two standard deviations from the result
$4.93(6) a_0^8 E_{\rm h} $ of~Ref.\cite{porsev}.

The detailed shape of the potential well usually gives unimportant
corrections to cold collision observables to the extent that the
scattering lengths and the long-range parameters~(\ref{bestfit}) are kept
fixed, see~\cite{gao}. However, sample calculations with modified inner
potentials show that to the current level of precision such corrections
are not fully negligible, and could be approximately accounted for
by multiplying by an extra factor of two the standard deviations in
Eq.~(\ref{bestfit}). With this proviso, in the following we provide
error bars as obtained from our current model that is thereby supposed
to give a sufficiently accurate description of the short range dynamics.

One should note that the potential parameters are statistically correlated.
For instance, if $C_6$ and $C_8$ were kept constant the position of a
given experimental feature could be approximately obtained by increasing
$a_S$ (i.e. by making the $^1\Sigma$ {\it less} binding) and concurrently
decreasing $a_T$ (i.e. by making the $^3\Sigma$ {\it more} binding).

As all parameters are left to vary correlations become more complex and
can be summarized for a linearized model in the symmetric covariance matrix:
\be
\cal{C}({\rm a.u.})=\left(
     \begin{array}{clrr}
        0.14 & 2.4[-2]& -0.47& -9.2[2]  \\
        \vdots     & 0.18 & -0.10 & 8.3[2] \\
             &  & 2.1 & 4.5[3] \\
             &  & \dotfill & 1.6[7]  \\
     \end{array}
  \right)
\ee
The $\cal C$ matrix has been used to compute error bars on the theoretical
resonance positions (second column in Tabs.~\ref{tab2},\ref{labtb1})
using standard error propagation whereas neglect of correlations might
lead to grossly overestimated uncertainties.

\begin{figure}[ht]
\vskip 12 pt
\includegraphics[width=\columnwidth,clip]{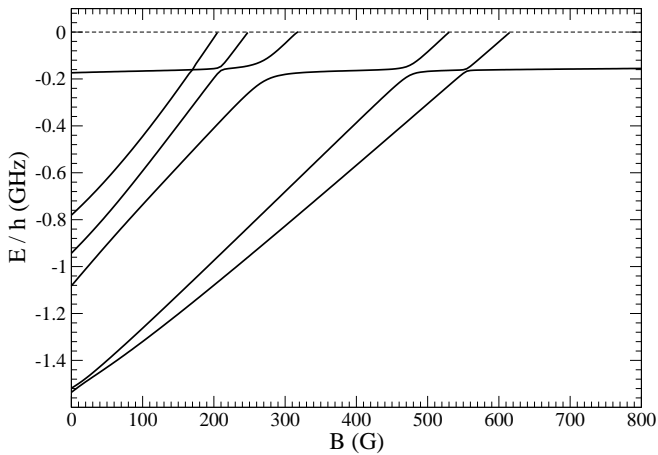}
\caption{Near threshold $^{39}$K$^{87}$Rb $\ell=0$ molecular
levels magnetically coupled
to atoms in the lowest Zeeman sublevel. The energy of
the separated atoms is taken as zero (dashed line).
Potential parameters are from Eq.~\ref{bestfit}.
 }
\label{fig3987mol}
\end{figure}

Our improved model is now used to determine the evolution of
molecular levels near dissociation, taking advantage of the
profound relation between near-threshold bound states and scattering
properties, see e.g.~\cite{gao}.
We first focus on the boson boson system for the experimentally relevant
case of $\ell=0$ molecules that can be magnetically associated starting
from atoms in the lowest Zeeman sublevel $|1 1 \rangle + | 1 1 \rangle $.

Levels closest to dissociation are usually strongly affected
by hyperfine interactions. This is especially the case when $a_S$
and $a_T$ have similar values since the splitting between singlet and
triplet vibrational levels is then small and the corresponding quantum
states are efficiently mixed by hyperfine couplings. Focusing on weak
magnetic fields, in this situation the diatomic is characterized by
Hund's case (e) quantum numbers $(f_a,f_b,f)$ where the total hyperfine
spin vector $\vec{f}=\vec{f}_a+\vec{f}_b$ is nearly exactly conserved.
As levels become more strongly bound the spin-exchange interaction first
causes decoupling of electron $\vec{s}_a$ and nuclear $\vec{\imath}_a$ spin
on the atom which has a smaller hyperfine coupling, potassium in the
present case. The hyperfine angular momentum of rubidium $\vec{f}_b$
then recouples with $\vec{s}_a$ because of spin-exchange forces to
form a new vector quantity which we denote here as $\vec{t}=\vec{s}_a +
\vec{f}_b$. Finally, the relatively weak potassium hyperfine interaction
forces $\vec{t}$ and $\vec{\imath}_a$ to form the total hyperfine angular
momentum $\vec{f}=\vec{t}+\vec{\imath}_a$. In this situation, the molecule is
represented by $\{ t f_b  f \}$ quantum numbers. 
Note that we have associated different brackets to different Hund's cases
in order to help to identify them. 
\begin{figure}[ht]
\vskip 12 pt
\includegraphics[width=\columnwidth,clip]{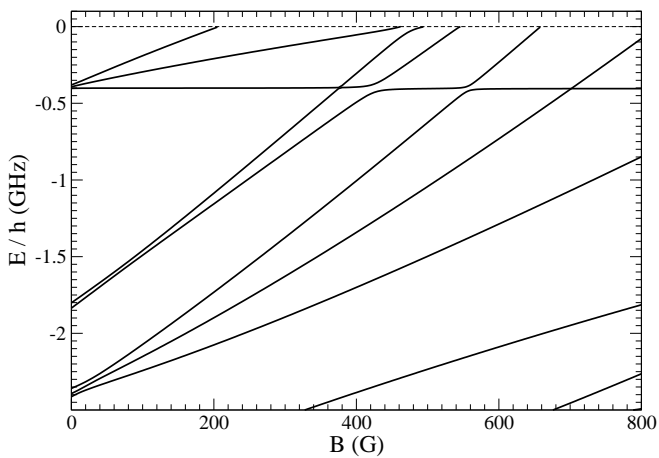}
\caption{
Same as Fig.~\ref{figlev1} but for $^{40}$K$^{87}$Rb.
 }
\label{figlev1}
\end{figure}

Molecular levels for $\ell=0$ molecules in the boson boson system
are presented in Fig.~\ref{fig3987mol}. The bound level at $-0.2$~Ghz
running parallel to the energy of the separated atoms is associated
to background scattering. That is, its position would correspond to
single channel scattering with the same background scattering length
and long-range coefficients. This is the only level characterized by
Hund's case (e) quantum numbers $(1,1,2)$. The five levels below,
four of which experimentally observed as resonances are all described
by the intermediate $\{ t f_b  f \}$ quantum numbers, see Tab.~\ref{tab2}.
One should note that near a resonance the molecular closed state channel
mixes with the open background channel~\cite{goral}. The size of the
region where this happens can be estimated as~\cite{goral}
\be
\frac{B-B_0}{\Delta} \ll
\frac{2 \mu a_{\rm bg}^2 s \Delta}{2 \hbar^2}, \label{broad}
\ee
where $\mu$ is the reduced mass and
\be
s=-\frac{\partial E}{\partial B}
\ee
is the magnetic moment of the molecule relative to that of the separated
atoms.
Using the parameters in Tab.~\ref{tab2} one can easily check that
$^{39}$K-$^{87}$Rb resonances are essentially closed-channel dominated as
according to Eq.~(\ref{broad}) mixing with the open channel is small over
most of the magnetic width $\Delta$. In this situation, near-resonance
effective models should involve at least on two channels characterized
by the parameters of Tab.~\ref{tab2}, see ~\cite{goral}.

We now discuss the fermion boson system. As the hyperfine splitting of
$^{40}$K is larger than the one of $^{39}$K and $a_S$ and $a_T$ have in
this case similar values, all resonances reported in Tab.~\ref{labtb1}
belong to Hund's case (e). That is, the exchange interaction is not strong
enough to decouple nuclear and electron spin of either atom. Molecular
levels for the Fermi Bose system are shown in Fig.~\ref{figlev1},
see also Ref.~\cite{KRblast} for similar results. Approximate values
of quantum numbers are found in Tab.~\ref{labtb1}. Finally, using the
parameters of Tab.~\ref{labtb1} and Eq.~(\ref{broad}) one finds that
fermion boson resonances range from closed channel dominated to an
intermediate situation, in which closed and open channel are mixed over
a significative fraction of the magnetic width.

So far in our procedure we have used the same interatomic potential
for the two isotopes thus assuming validity of the Born-Oppenheimer
approximation. In order to quantify possible breakdown effects \cite{Tie1}
we fit our data by varying {\it independently} the short range potential
for the two isotopes. Result of the fit is then
\bea
&& a_S(40-87)=-110.8 \, a_0  \nonumber \\
&& a_T(40-87)=-213.8 \, a_0 \nonumber \\
&& a_S(39-87)=1.98~10^3 \, a_0 \nonumber \\
&& a_T(39-87)=35.6 \, a_0 \nonumber \\
&& C_6=4291 \, a_0^6E_h \nonumber \\
&& C_8=4.80~10^5 \, a_0^8E_h \nonumber ,
\eea
corresponding to  $\chi^2=0.93$, a slightly larger value than the one
found above due to the diminished number of degrees of freedom.
Note that these best fit parameters are fully consistent with the values
obtained assuming mass scaling.
\begin{table}[tbp]
\begin{center} \caption{Singlet and triplet $s$-wave scattering lengths for collisions
between K and Rb isotopic pairs based on our spectroscopic data for both the Fermi Bose and the
Bose Bose mixture.} \vskip 12pt
\begin{tabular}{c | c c }
\hline \hline
 K-Rb pair & $a_S(a_0)$ & $a_T(a_0)$ \\
\hline
 39-85 & 33.78(6)  & 63.27(2) \\
 39-87 & 1.98(4)\,$10^3$ & 35.61(3) \\
 40-85 & 65.39(5) & -28.63(6) \\
 40-87 & -110.6(4) & -214.0(4) \\
 41-85 & 103.25(6) & 349.0(4) \\
 41-87 & 7.13(9) & 163.82(6) \\
 \hline \hline
\end{tabular}
\end{center}
\label{tab3}
\end{table}

\begin{figure}[ht]
\vskip 12 pt
\includegraphics[width=\columnwidth,clip]{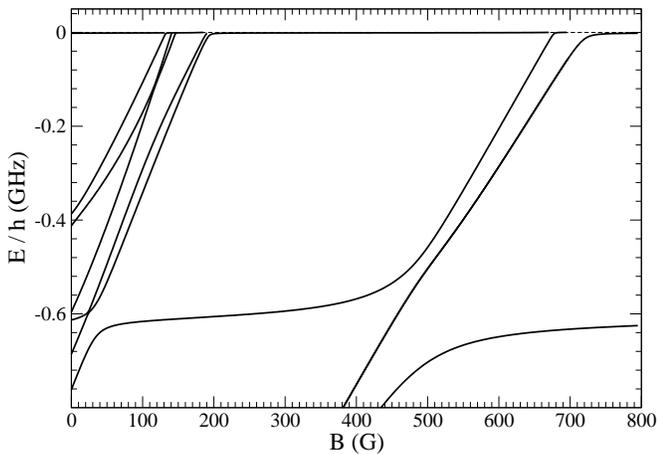}
\caption{Same as Fig.~\ref{fig3987mol} but for $^{41}$K$^{85}$Rb.
}
\label{fig4185}
\end{figure}
We also remark that theoretical resonance positions do not show
any preferential, positive or negative shift with respect to the
experimental ones. We can conclude that even at the present level of
precision no evidence is found for breakdown of the Born-Oppenheimer
approximation. Mass-scaling can then be used for predicting properties of
other isotopes. In particular, the $a_{S,T}$ along with the long-range
coefficients determined in this work are sufficient in order to predict
all relevant threshold properties of any K-Rb pair.

\begin{figure}[ht]
\vskip 12 pt
\includegraphics[width=\columnwidth,clip]{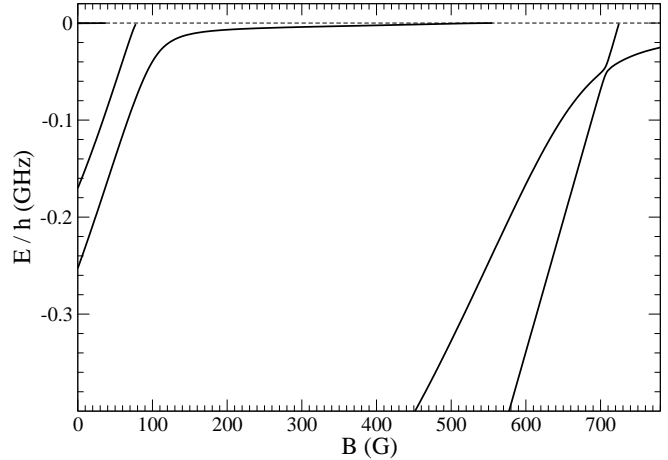}
\caption{Same as Fig.~\ref{fig3987mol} but for $^{41}$K$^{87}$Rb.
}
\label{fig4187}
\end{figure}

\begin{table*}[th]
\begin{center}
\caption{Predicted zero-field $s$-wave scattering lengths $a$ for the
absolute ground state of K-Rb isotopes.
Positions $B_{\rm th}$, widths $\Delta_{\rm th}$, background scattering lengths $a_{\rm bg}$, and
magnetic moments are also provided for Feshbach resonances of two isotopic
pairs of main experimental interest.
} \label{table3} \vskip 12pt
\begin{tabular*}{10cm}{@{\extracolsep{\fill}}l@{\hspace{0.5cm}}|c@{\hspace{0.5cm}}cccc}
\hline \hline
K-Rb & $a$ ($a_0$) & $B_{\rm th}$~(G)&$ \Delta_{\rm th}$~(G)& $a_{\rm bg}$ ($a_0$)& $-s$ ($\mu_B$) \\
\hline
39-85&$58.01(2)$&&&\\
\hline
40-85&$-21.06(6)$&&&\\
\hline
39-87&$28.29(3)$&&&\\
\hline
40-87&$-184.4(3)$&&&\\
\hline
41-85&$283.1(3)$&$132.39(7)$&0.19&242&2.33 \\   
&&$140.98(5)$&$2.0 ~10^{-4}$&242&3.42\\     
&&$146.4(3)$&0.025&242&2.88\\     
&&$185.2(9)$&3.5 & 327 & 2.14\\     
&&$191.72(7)$&0.48 & 327 & 2.14 \\     
&&$672.19(15)$&5.7&343 & 1.89 \\   
&&$695.90(12)$&14&343 & 1.70   \\    
\hline
41-87&$640(3)$&$39.4(2)$&37&284 & 1.65  \\           
&&$78.92(9)$& 1.2 & 284&     1.59      \\           
&&$558.0(4)$& 81  &  173&     1.14     \\            
&&$724.8(3) $&0.07& 90 &      1.93     \\            
 \hline \hline
\end{tabular*}
\end{center}
\end{table*}

The $s$-wave singlet and triplet scattering lengths are shown in
Tab.~\ref{tab3} for all isotopic combinations. Both $a_S$ and $a_T$
are consistent with our previous determination~\cite{ferlaino} if one
corrects for the different number of bound states supported by the
singlet potential, as specified in that work. Our values are consistent
with the results of Pashov {\it et al.}~\cite{pashov} if one assumes
for that work the same error bars of Ref.~\cite{ferlaino}.

We also predict with high precision the value of the $s$-wave scattering
length $a$ for the absolute ground state, see Tab.~\ref{table3}. For two
especially interesting pairs discussed in Ref.~\cite{ferlaino},
the position of magnetic Feshbach resonances is recalculated with
the current parameters \cite{minardi}. Positions are slightly shifted with respect
to the ones of \cite{ferlaino} because of the different number of
bound states in the singlet potential. They seem consistent 
with the plots shown in Ref.~\cite{KRblast} which does not
otherwise provide numerical values to compare with. We refer the reader
to Ref.~\cite{ferlaino} for a discussion of possible applications of
resonances in these specific K-Rb isotopic systems.

Finally, we present in Figs.~\ref{fig4185} and \ref{fig4187} near
threshold-molecular potentials for the two pairs. One may note a level
very close to dissociation associated to background scattering for
both isotopic combinations, corresponding to a large positive $a_{\rm
bg}$. Avoided crossings of such bakground level with magnetic field
dependent molecular levels give rise to strong resonance shifts that
can given a simple analytic expression in terms of $a_{\rm bg}$, $C_6$
and $\Delta$~\cite{goral}. For instance, the resonance at 39~G in
Fig.~\ref{fig4187} arises from the molecular level occurring at zero
field near -0.25~GHz.

Again, using the parameters of Tab.~\ref{table3} one can see that
both isotopic combinations present both broad open channel dominated
resonances, which can be modelled theoretically by a single effective
channel, and narrow closed channel dominated ones. Availability of
such a broad range of properties should pave the way to the exploration
of different quantum regimes in ultracold binary gases. Our data also
provide a needed piece of information for the calculation of Franck-Condon
overlap matrix with electronically excited states and for implementing
efficient transfer scheme to low vibrational levels using Feshbach
molecules as a bridge.

\section{Conclusions and outlook}
\label{sec4}
In conclusion, we have performed extensive Feshbach spectroscopy
of an ultracold $^{39}$K-$^{87}$Rb mixture. Combination of new
spectroscopic measurements on this system with data relative to
the Fermi Bose $^{40}$K-$^{87}$Rb system has allowed us to improve
significantly the accuracy of our model. Interisotope analysis 
determines near-threshold parameters with better precision and
fixes the number of bound levels supported by the interaction
potentials. To the present level of precision no evidence for breakdown
of the Born-Oppenheimer approximation has been found. Therefore, we
have determined by a straightforward mass scaling procedure different
scattering properties for all K-Rb isotopic mixtures. The present results
combined with information on short range potentials is of crucial
importance in order to determine the most convenient strategy for
association of weakly bound molecules and their optical transfer into
deeper bound states.


\begin{thebibliography}{99}
\bibitem{feshbach} E. Tiesinga, B. J. Verhaar, H. T. C. Stoof, Phys.
Rev. A {\bf 47}, 4114 (1993).
\bibitem{sodium} S. Inoyue, M. R. Andrews, J. Stenger, H.-J. Miesner, D.M. Stamper-Kurn, and W.
Ketterle, Nature {\bf 392}, 151 (1998).
\bibitem{Feshmag} J. Stenger, S. Inouye, M. R. Andrews, H. J. Miesner, D. M. Stamper-Kurn, and W. Ketterle,
Phys. Rev. Lett. {\bf 82}, 2422 (1999);
\bibitem{fermioni} T. Loftus, C. A. Regal, C. Ticknor, J. L. Bohn, and D. S. Jin, Phys. Rev. Lett. {\bf 88}, 173201 (2002).
\bibitem{mix} S. Inouye, J. Goldwin, M. L. Olsen, C. Ticknor, J. L. Bohn, and D. S. Jin, Phys. Rev. Lett. {\bf 93}, 183201 (2004).
\bibitem{zaccanti} M. Zaccanti, C. D'Errico, F. Ferlaino, G. Roati, M. Inguscio, and G. Modugno,
Phys. Rev. A 74, 041605(R) (2006).
\bibitem{bongs} S. Ospelkaus, C. Ospelkaus, L. Humbert, K. Sengstock, and K. Bongs,
Phys. Rev. Lett. {\bf 97}, 120403 (2006).
\bibitem{molecular gases} S. Jochim, M. Bartenstein, A. Altmeyer, G. Hendl, S. Riedl, C.
Chin, J. Hecker Denschlag and R. Grimm, Science {\bf 302}, 2101
(2003)\textit; M. Greiner, C.A. Regal, and D.S. Jin, Nature {\bf
426}, 537 (2003); M. W. Zwierlein, C. A. Stan, C. H. Schunck, S. M.
F. Raupach, S. Gupta, Z. Hadzibabic, and W. Ketterle, Phys. Rev.
Lett. {\bf 91}, 250401 (2003).
\bibitem{eteromol} C. Ospelkaus, S. Ospelkaus, L. Humbert, P. Ernst, K. Sengstock, and K. Bongs, Phys. Rev. Lett. {\bf 97}, 120402 (2006); S. B. Papp and C. E. Wieman, Phys. Rev. Lett. {\bf 97}, 180404 (2006); J. J. Zirbel, K.-K. Ni, S. Ospelkaus, J. P. D'Incao, C. E. Wieman, J. Ye, D. S. Jin . arXiv:0710.2479; J. J. Zirbel, K.-K. Ni, S. Ospelkaus, T. L. Nicholson, M. L. Olsen, C. E. Wieman, J. Ye, D. S. Jin, P. S. Julienne, arXiv:0712.3889.
\bibitem{rempe} A. Marte, T. Volz, J. Schuster, S. Durr, G. Rempe, E. G. M. van Kempen, and B. J. Verhaar
Phys. Rev. Lett. 89, 283202 (2002)
\bibitem{Chu} C. Chin, V. Vuletic, A. J. Kerman, and S. Chu, Phys. Rev. Lett. \textbf{85}, 2717 (2000)
\bibitem{modugno} G. Modugno, G. Ferrari, G. Roati, R. J. Brecha, A. Simoni, and M. Inguscio,
 Science {\bf 294} 1320 (2001)
\bibitem{roati2}G. Roati, F. Riboli, G. Modugno, and M.
Inguscio, Phys. Rev. Lett. {\bf 89}, 150403 (2002).
\bibitem{roati07} G. Roati, M. Zaccanti, C. D'Errico, J. Catani, M. Modugno, A. Simoni, M. Inguscio, and G. Modugno, Phys. Rev. Lett. {\bf99}, 010403 (2007).
\bibitem{desarlo} L. De Sarlo, P. Maioli, G. Barontini, J. Catani, F. Minardi, and M.  Inguscio, Phys. Rev. A {\bf 75}, 022715 (2007).
\bibitem{ferlaino} F. Ferlaino, C. D'Errico, G. Roati, M. Zaccanti, M. Inguscio, G. Modugno, A.
Simoni, Phys. Rev. A  {\bf 73}, 040702(R) (2006).
\bibitem{KRbJila} S. Inouye, J. Goldwin, M. L. Olsen, C. Ticknor, J. L. Bohn, and D. S. Jin,
Phys. Rev. Lett. {\bf 93}, 183201 (2004).
\bibitem{KRblast} C. Klempt, T. Henninger, O. Topic, J. Will, W. Ertmer, E. Tiemann, and J. Arlt,
Phys. Rev. A {\bf76}, 020701(R) (2007) .
\bibitem{kotochigova} S. Kotochigova, P. S. Julienne, and E. Tiesinga,
Phys. Rev. A {\bf68}, 022501 (2003)
\bibitem{stwalley} D. Wang, J. Qi, M. F. Stone, O. Nikolayeva, H. Wang, B. Hattaway, S. D. Gensemer, P. L. Gould, E. E. Eyler, and W. C. Stwalley, Phys. Rev. Lett. {\bf93}, 243005 (2004).
\bibitem{pashov} A. Pashov, O. Docenko, M. Tamanis, R. Ferber, H. Knoeckel, and E. Tiemann,
Phys. Rev. A {\bf76}, 022511 (2007)
\bibitem{pavel} P. Sold\'an and V. {\v S}pirko, J. Chem. Phys. {\bf 127}, 121101 (2007).
\bibitem{zemke} W. T. Zemke, R. Cot\'e, and W. C. Stwalley, Phys. Rev. A {\bf 71}, 062706 (2005)
\bibitem{Esry} J. P. D'Incao and B. D. Esry, Phys. Rev. Lett. {\bf 94}, 213201 (2005).
\bibitem{bohn} C. Ticknor, C. A. Regal, D. S. Jin, and J. L. Bohn, Phys. Rev. A {\bf 69}, 042712 (2004).
\bibitem{amiot} C. Amiot and J. Verg{\`e}s, J. Chem. Phys. {\bf 112}, 7068 (2000).
\bibitem{rousseau} S. Rousseau, A. R. Allouche, and M. Aubert-Fr{\'e}con,
J. Mol. Spectrosc. {\bf 203}, 235 (2000).
\bibitem{rkr1} R. J. Le Roy, {\bf RKR1 2.0}: {\it A Computer Program Implementing
the First-Order RKR Method for Determining Diatomic Molecule Potential Energy Curves},
University of Waterloo Chemical Physics Research Report CP-657R (2004).
\bibitem{Mies} F. H. Mies, C. J. Williams, and P. S. Julienne, J. Res. Natl. Inst. Stand. Technol. {\bf 101}, 521 (1996).
\bibitem{private} P. S. Julienne and S. Kotochikova, {\it private communication}.
\bibitem{KNJP} C. D'Errico, M. Zaccanti, M. Fattori, G. Roati, M. Inguscio, G. Modugno,
and A. Simoni, New J. Phys {\bf 9}, 223 (2007).
\bibitem{derevianko} A. Derevianko, J. F. Babb, and A. Dalgarno, Phys. Rev. A {\bf 63}, 052704 (2001).
\bibitem{porsev} S. G. Porsev and A. Derevianko, J. Chem. Phys. {\bf 119}, 884 (2003).
\bibitem{gao} Bo Gao, E. Tiesinga, C. J. Williams, and P. S. Julienne, Phys. Rev. A { \bf 72}, 042719 (2005).
\bibitem{goral} K. Goral, T. Koehler, S. A. Gardiner, E. Tiesinga, and P. S. Julienne, J. Phys. B {\bf 37}, 3457 (2004).
\bibitem{Tie1} S. Falke, E. Tiemann, and C. Lisdat, arXiv:0.706.2290v1.
\bibitem{minardi} Recently, few of the predicted features for the $^{41}$K-$^{87}$Rb isotopic pair have been experimentally observed at the predicted magnetic field position. F. Minardi, {\it private communication}.
\end{thebibliography}
\end{document}